# The time resolved measurement of ultrashort THz-band electric fields without an ultrashort probe

D. A. Walsh[1,a], E. W. Snedden[1], S. P. Jamison[1,2]

[1]Accelerator Science and Technology Centre, STFC Daresbury National Laboratory, Warrington, WA4 4AD, United Kingdom

[2]Photon Science Institute, The University of Manchester, Manchester. M13 9PL, United Kingdom

The time-resolved detection of ultrashort pulsed THz-band electric field temporal profiles without an ultrashort laser probe is demonstrated. A non-linear interaction between a narrow-bandwidth optical probe and the THz pulse transposes the THz spectral intensity and phase information to the optical region, thereby generating an optical pulse whose temporal electric field envelope replicates the temporal profile of the real THz electric field. This optical envelope is characterised via an autocorrelation based FROG measurement, hence revealing the THz temporal profile. The combination of a narrow-bandwidth, long duration, optical probe and self-referenced FROG makes the technique inherently immune to timing jitter between the optical probe and THz pulse, and may find particular application where the THz field is not initially generated via ultrashort laser methods, such as the measurement of longitudinal electron bunch profiles in particle accelerators.

In applications such as terahertz time domain spectroscopy (THz-TDS) the characterisation of ultrashort pulse THz radiation has commonly proceeded through the sampling of the THz field with an optical probe of even shorter duration in an electro-optically responsive material [1-3]. This process of electro-optic sampling (EOS) naturally runs into difficulty when the THz pulse duration is comparable to or shorter than that of the available probe, thus limiting temporal resolution of the retrieved field and preventing an accurate measurement. A similar problem was historically encountered in the characterisation of ultrashort optical radiation, but the development of combined time- and frequency-domain methods such as FROG [4-6] and SPIDER [7] provided the means to recover the temporal envelope, and phase evolution, of such pulses. The absence of such self-referenced techniques at THz frequencies can be attributed in part to the lack of suitably high pulse energies and compatible non-linear materials. More fundamentally, in THz-TDS applications the carrier envelope phase (CEP), which is not retrieved in standard FROG methods, is an essential part of the electric field that needs to be characterised. While this CEP problem has recently been overcome in principle by "ReD-FROG" [8], here the FROG CEP ambiguity is bypassed in the reconstruction of the THz field through conversion of the full THz field, including the "carrier" and the envelope, into an optical envelope,.

In addition to the limitations to the temporal resolution achievable with a given probe pulse in THz-TDS, the method is subject to deficiencies in the presence of timing jitter between the probe and THz pulses. In best cases the timing jitter contributes to a blurring of the temporal information; in worst cases the delay mapping is completely lost and the measurement invalidated. There are a number of methods which have been developed to enable single-shot characterisation via EOS which remap the time delay spatially or spectrally. Spatial encoding is possible in both non-collinear [9] and pulse-front tilt schemes [10], but necessarily reduces the THz field as the pulse is spread out. Additional care must also be taken in that the handling of the beams and their spatial intensity profiles should not distort the temporal measurement [11]. Spectral encoding utilises a stretched ultrashort pulse to map time delay to frequency content and is subject to fundamental limitations in temporal resolution (of the order of hundreds of femtoseconds) in its most common form [12]. Enhanced methods to improve the temporal resolution using a cross-correlation [13] or cross-FROG [14] do exist, however, all these techniques still require relatively low levels of timing jitter as the temporal window of the sampling pulse is limited, and both still require ultrashort laser sources.

---

[a] Author to whom correspondence should be addressed. Electronic mail: david.walsh@stfc.ac.uk

In this letter we report the time resolved detection of 0-3.0 THz electric field waveforms using a narrow-bandwidth quasi-CW optical probe. The time resolution of the electric field measurement is no longer derived from the duration of the probe, and represents a method for achieving high time resolution measurements without the need for ultrashort lasers. The concept is demonstrated through a benchmarked proof-of-principle experiment involving the $\chi^{(2)}$ interaction between optical probe and THz waveforms in ZnTe. By using a narrow bandwidth optical probe the THz spectral information is remapped into up-converted optical waves, thereby reproducing the THz waveform in the envelope of an optical pulse. This optical envelope ($\equiv$ spectral amplitude and phase) is then characterised through self-referenced FROG allowing the THz pulse to be recovered.

The principle reported here allows for sub-100 fs resolution in THz waveform measurements with nanosecond or CW optical probes. For our proof-of-concept demonstration a 10 ps duration 'quasi-CW' probe is generated through 4-f spectral filtering of a 50 fs duration pulse. This allows the THz pulse to be independently characterised with conventional electro-optic sampling with 50 fs pulses, and for the THz pulse to be conveniently generated from the same laser.

The Electro-optic (EO) effect can be generally described as the sum- and difference-frequency generation between the frequency components of the optical probe and THz pulses [15, 16]. For the commonly used EO materials ZnTe and GaP the largest nonlinear effect is produced in (110) cut crystals where the optical and THz fields are polarised parallel to $[1\bar{1}0]$. The principal axes of the interaction in this system are then at ±45° to this polarisation [17-19]. Using the result of Jamison *et al*, the wave generated by the EO interaction in the 2 principle axes ($i$ = 1 or 2) of the material in the time domain is expressed as:

$$E_i(t;\tau) \propto \beta_i \frac{d}{dt}\left[E^{opt}(t).E^{THz}_{eff}(t-\tau)\right]$$

where $E^{opt}(t)$ denotes the input optical field, $E^{THz}_{eff}(t)$ is the effective input field including the effects of dispersion in $\chi^{(2)}$ and phase matching, and $\tau$ is the temporal delay between the optical and THz pulses. For the geometry specified above the axis dependent constant $\beta_i$ takes the value of +1 or -1 depending on the axis, $i$ = 1 or 2, being examined. As such, the newly generated waves can be separated from the input via a polariser orthogonal to the input optical polarisation. With dispersion data at THz frequencies available for common media determination of $E^{THz}(t)$ follows from knowledge of $E^{THz}_{eff}(t)$ [16, 20].

If a narrow bandwidth continuous wave or long pulse (i.e. one that does not change significantly on the timescale of the THz field) is used for the probe field, and the optical probe field oscillation is much faster than any temporal variations in the THz pulse, then this optical carrier frequency dominates the derivative whereas the optical envelope is constant. This results in the mapping of the THz profile into a new pulse envelope after the crossed polariser geometry according to:

$$E(t;\tau) \propto 2\left(\frac{d}{dt}E^{opt}(t)\right) \cdot E_{eff}^{THz}(t).$$

This process is analogous to musical transposition, where a piece of music is changed from one key to another: the notes (spectral amplitudes) are shifted by a fixed frequency and the relative note timings (spectral phases) are kept the same, thereby preserving the melody (temporal profile). Due to this similarity we call the process Electro-Optic Transposition (EOT). We note that at THz frequencies the absolute phase is essential to the reproduction of the pulse profile, but once shifted to the optical regime the absolute phase only affects the optical carrier wave; the optical envelope, and so the THz temporal profile, can be faithfully retrieved via standard FROG techniques thereby bypassing the CEP ambiguity.

An experimental validation of the EOT method was performed using the setup outlined in Figure (1). The THz source was a large area (75 mm x 75 mm), lt-GaAs, photo-conductive antenna (PCA) with a horizontally applied bias of ~100 kV. This was pumped with 90% of the output of a Ti:Sapphire regenerative amplifier (500 Hz, ~1.5 mJ, 50 fs transform-limited pulse duration). The remaining 10% input pulse energy was propagated through a scanning delay line and a zero-dispersion 4-f spectral filter to form the optical probe. The bandwidth of the optical probe could then be varied using slit placed in the Fourier plane of the 4-f filter.

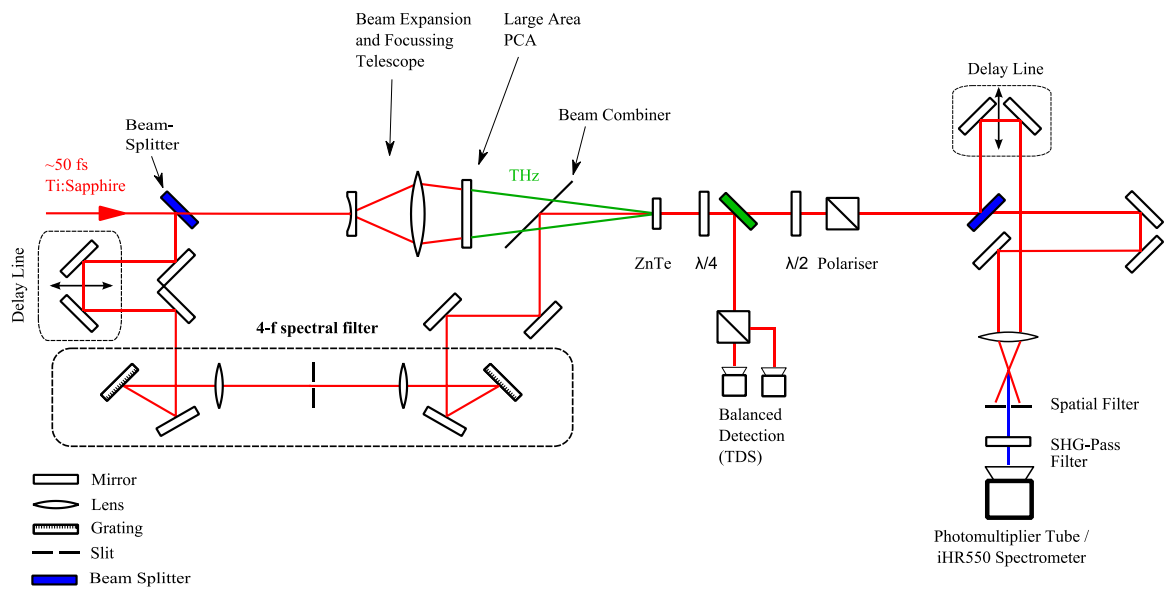

Figure (1) Experimental layout for the demonstration of EOT-FROG.

Both THz and optical probe were combined coaxially with the THz pulse using an ITO coated glass substrate and directed on to a 4 mm ZnTe crystal. A flip mirror was used to direct the ZnTe output to either a balanced detection set-up for EOS, or to a crossed polarisation analyser for EOT. A quarter wave plate after the ZnTe was implemented differently in each scheme: circularising the polarisation of the transmitted light in balanced detection, and correcting for any impurity/stress-induced residual birefringence in the ZnTe in EOT.

Measurement in the balanced detection scheme was performed using lock-in amplifier techniques referenced to the 250Hz pulsing of the PCA bias voltage; the resulting time domain sampling of the THz waveform, using a 50 fs probe pulse, is shown in Figure 3. For EOT measurements the slit in the 4-f filter was set to produce probe pulses 10 ps in duration of which a small amount was leaked through the crossed polariser geometry and coupled into a bespoke non-collinear auto correlator based on 100 μm thick BBO cut for second harmonic generation of 800 nm light. This allowed confirmation of the probe duration, and the signal from this was aligned into an iHR550 imaging spectrometer coupled to a DiCAM pro intensified CCD camera, completing the FROG arrangement. After alignment the probe was extinguished again in the crossed polarised arrangement to facilitate measurement of the transposed pulses alone. Due to the low energy that was available in the transposed pulse (<0.5 nJ) a multiple shot averaging of the intensified camera images was required to measure the SHG spectrum at each delay point. This involved 256 frame integrations on the iCCD before readout, and then summing 30 of these samples. The complete FROG spectrograms were recorded over 128 points with a step size of 62.5 fs.

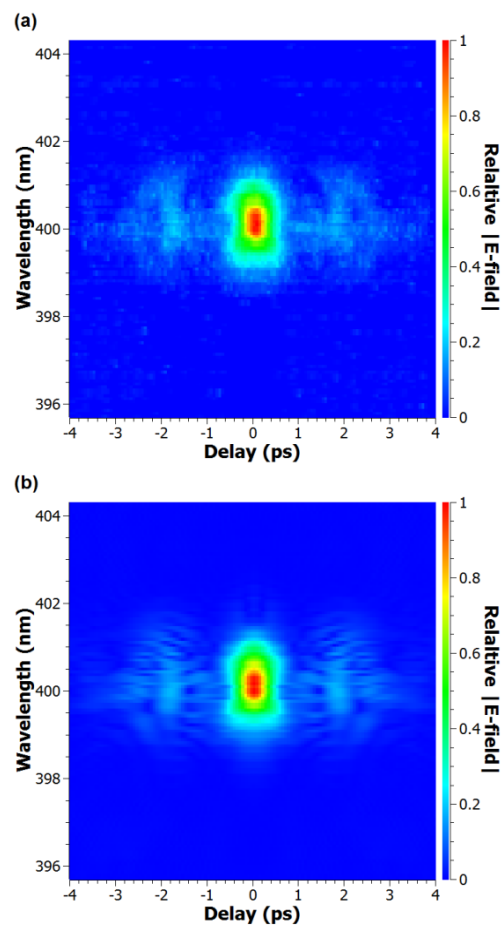

Figure (2) (a) The measured spectrogram after noise reduction and (b) the recovered spectrogram. Both are displayed in false colour and proportional to the electric field envelope magnitude to enhance visibility.

The FROG spectrogram was analysed using Femtosoft Technologies commercial FROG retrieval code "FROG3", including image pre-processing for noise and background. The (noise-filtered) measured spectrogram is shown alongside the recovered spectrogram in Figure 2. The recovered spectrogram is in good agreement with the measured one, despite the remaining noise, yielding a FROG error of 0.007 for the 128x128 spectrogram.

The temporal intensity envelope and phase of the EOS and recovered EOT signals are compared in Figure (3). Despite the high level of noise in the measured spectrogram it is quite clear that the temporal profile has been recovered quite faithfully.

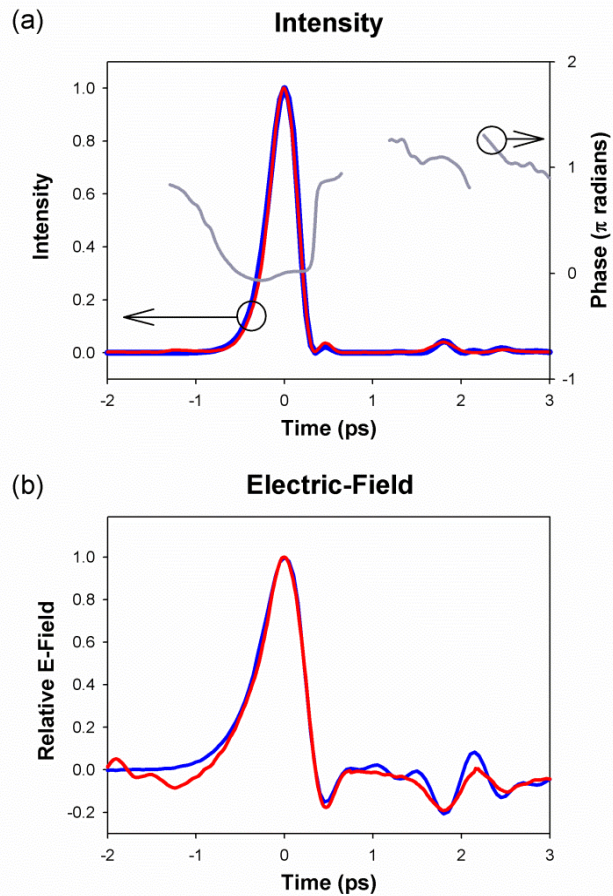

Figure (3) (a) The |E-field|² of the THz-TDS trace in blue overlaid with the FROG-recovered intensity envelope in red. The grey dashed line represents the FROG-retrieved relative phase. (b) The E-field recovered via THz-TDS.

For the data set presented here, the limited signal to noise together with the well known SHG FROG ambiguity for well-separated pulses, led to the trailing oscillation at 2 ps having a phase ambiguity of $\pm n\pi$, where $n$ is an integer. In this case we have manually subracted $\pi$ in order for the sign of the recovered THz field to match. Note that the polarity shift occuring within a "continuous" pulse at 0.5 ps was correctly retreived from the FROG recovered phase. It is envisaged that in many situations a higher signal to noise ratio would mitigate this ambiguity as the interaction at intermediate delays between the pulses would be measurable.

In summary, we have reported an experimental scheme that will enable the time resolved measurement of ultrashort THz radiation without the need for an ultrashort optical probe pulse of duration much less than the desired time resolution. This demonstration used 10 ps long probe pulses, but the principle is easily extended to the use of nanosecond long pulses for which the jitter tolerance between the THz pulse and that of the laser increases by a further two orders of magnitude. The only constraints on the EOT method are that the two pulses temporally overlap, and that any temporal variation in the THz and optical envelopes can be considered negligible to the optical carrier frequency. There remains the practical issue of having sufficient pulse energy to perform the FROG measurement, but these may be overcome via the implementation of a phase preserving amplification technique for the EOT pulse such as optical parametric chirped pulse amplification [21], and/or a more sensitive FROG arrangement such as GRENOUILLE. Moving to GRENOUILLE will also enable single shot measurements of the temporal profile to be made with higher resolution than is possible via other single shot methods, being limited by the EO material response rather than the probe pulse properties.